\documentclass[draftcls, onecolumn, 10pt]{IEEEtran}
\usepackage{amssymb,amsmath}
\usepackage[dvips]{graphicx}
\usepackage{amsfonts}
\usepackage{subfigure}
\usepackage{booktabs,multirow}
\usepackage{color}
\usepackage{cite}
\usepackage[table]{xcolor}

\IEEEoverridecommandlockouts

\begin{document}

\title{5G Ultra-Dense Cellular Networks}

\author{\normalsize
Xiaohu Ge$^1$,~\IEEEmembership{Senior~Member,~IEEE,} Song Tu$^1$, Guoqiang Mao$^2$,~\IEEEmembership{Senior~Member,~IEEE}, Cheng-Xiang Wang$^3$,~\IEEEmembership{Senior~Member,~IEEE}, Tao Han$^1$,~\IEEEmembership{Member,~IEEE}\\
\vspace{0.70cm}
\small{
$^1$School of Electronic Information and Communications\\
Huazhong University of Science and Technology, Wuhan 430074, Hubei, P. R. China.\\
Email: \{xhge, u201013039, hantao\}@mail.hust.edu.cn\\
\vspace{0.2cm}
$^2$School of Computing and Communications\\
University of Technology Sydney, Australia.\\
Email: g.mao@ieee.org\\
\vspace{0.2cm}
$^3$Institute of Sensors, Signals and Systems, \\
School of Engineering \& Physical Sciences,
Heriot-Watt University, Edinburgh, EH14 4AS, UK.\\
Email: cheng-xiang.wang@hw.ac.uk}\\
\thanks{\small{ Submitted to IEEE Wireless Communications.}}
}

\renewcommand{\baselinestretch}{1.2}
\thispagestyle{empty}
\maketitle
\thispagestyle{empty}
\setcounter{page}{1}\begin{abstract}
 Traditional ultra-dense wireless networks are recommended as a complement for cellular networks and are deployed in partial areas, such as hotspot and indoor scenarios. Based on the massive multiple-input multi-output (MIMO) antennas and the millimeter wave communication technologies, the 5G ultra-dense cellular network is proposed to deploy in overall cellular scenarios. Moreover, a distribution network architecture is presented for 5G ultra-dense cellular networks. Furthermore, the backhaul network capacity and the backhaul energy efficiency of ultra-dense cellular networks are investigated to answer an important question, \emph{i.e., how much densification can be deployed for 5G ultra-dense cellular networks}. Simulation results reveal that there exist densification limits for 5G ultra-dense cellualr networks with backhaul network capacity and backhaul energy efficiency constraints.

\end{abstract}

\IEEEpeerreviewmaketitle

\newpage
\section{Introduction}

To meet 1000$\times $ wireless traffic volume increment in the next decade, the fifth generation (5G) cellular network is becoming a hot research topic in telecommunication industries and academics. Firstly, the massive multiple-input multi-output (MIMO) technology was proposed to improve the spectrum efficiency of 5G mobile communication systems \cite{Hoydis2013}. Secondly, the millimeter wave communications was presented to extend the transmission bandwidth for 5G mobile communication systems \cite{Rappaport}. Furthermore, the small cell concept has been appeared to raise the throughput and save the energy consumption in cellular scenarios \cite{Wang2014}. To satisfy the seamless coverage, a larger number of small cells have to be densely deployed for 5G cellular networks. As a consequence, the ultra-dense cellular network is emerging as one of core characteristics for 5G cellular networks. However, the study of ultra-dense cellular networks is still in an initial stage. Some basic studies, such as the network architecture and cellular densification limits need to be more investigated for future 5G cellular networks.

In the third generation (3G) cellular networks, the densification of macrocell base stations (BSs) aims to improve the transmission rate in partial areas, such as macrocell BSs deployed in urban areas. To avoid interference in adjacent macrocell BSs, the frequency reuse and sectorized BS technologies have been developed for macrocell densification, where the density of macrocell BSs is about 4-5 BS/km$^2$. In the fourth generation (4G) cellular networks, such as Long Term Evolution-Advanced (LTE-A) mobile communication systems, the microcell BSs, e.g., hotspot BSs and femtocell BSs, have been deployed to satisfy the high speed transmission in specified regions, where the density of microcell BSs is approximate 8-10 BS/km$^2$. Moreover, all above BSs are directly connected by gateways and all backhaul traffic is forwarded by fiber links or broadband Internet. In 3G and 4G cellular networks, the aim of BSs densification is to improve the wireless transmission rate in partial regions and the most challenge of BSs densification is the interference coordination for cellular networks.
In 5G cellular networks, the massive MIMO antennas will be integrated into BSs, where hundreds antennas are utilized for transmitting Gbits level wireless traffic. When the 5G BS transmission power is constrained at the same level of 4G BS transmission power, every antenna transmission power at 5G BS has to be decreased 10-20 times compared with every antenna transmission power at 4G BS. As a consequence, the radius of 5G BS has to be decreased one magnitude considering the decrease of transmission power at every antenna. Another potential key technology for 5G cellular networks is the millimeter wave communication technology, which is expected to provide hundreds MHz bandwidth for wireless transmissions. However, the transmission distance of millimeter wave communications has to be restricted into 100 meters considering the propagation degradation of millimeter wave in the atmosphere. Motived by above two technologies, small cell networks have been presented for 5G cellular networks. To satisfy the seamless coverage, the density of 5G BS is highly anticipated to come up to 40-50 BS/km$^2$. Therefore, the future 5G cellular network is an ultra-dense cellular network.

Some initial studies involved with ultra-dense wireless networks were explored in \cite{Yunas,Soret,Asadi,Bhushan,Condoluci,Ge2014,He2007,Hur2013,Dehos}. Yunas \emph{et al.} investigated the spectrum and energy efficiency of ultra-dense wireless networks under different deployment strategies, such as the densification of classical macrocell BSs, ultra-dense indoor femtocell BSs and outdoor distributed antenna systems \cite{Yunas}. Soret \emph{et al.} discussed the interference problem for dense scenarios of LTE-A cellular networks and proposed two algorithms to apply time domain and frequency domain small cell interference coordination for dense wireless networks \cite{Soret}. Based on LTE and WiFi technologies, a joint coordinated intra-cell and inter-cell resource allocation mechanism was proposed to opportunistically exploit network density as a resource \cite{Asadi}. However, these solutions were mainly presented for 4G cellular networks, such as LTE networks. Bhushan \emph{et al.} discussed advantages of network densification, which has included the spatial densification, e.g., dense deployment of small cell and spectrum aggregation, i.e., utilizing larger portions of radio spectrum in diverse bands for 5G networks \cite{Bhushan}. Moreover, in this densification network architecture, the dense deployment of small cell is limited in indoor scenarios and users in outdoor scenarios are still covered by traditional macrocells. By absorbing the machine-type communication (MTC) traffic via home evolved NodeBs, a new architecture was proposed by the use of small cells to handle the massive and dense MTC rollout \cite{Condoluci}. As concluded in \cite{Bhushan,Condoluci}, these dense wireless networks are the complement for existing macrocell networks.
Considering the backhaul traffic challenge in 5G small cell networks, the central and distributed wireless backhaul network architectures were compared in \cite{Ge2014}. Simulation results suggested that the distributed wireless backhaul network architecture is more suitable for future 5G networks employing massive MIMO antennas and millimeter wave communication technologies. It is noteworthy that the distributed wireless backhaul network architecture has also been discussed for IEEE 802.16 mesh networks in \cite{He2007}. Considering that the radius of IEEE 802.16 BSs is typically 1500 meters, which is much larger than the 50-100 m radius of small cells, IEEE 802.16 mesh networks are not ultra-dense wireless networks. Therefore, the small-cell density deployment bottleneck is not a problem for IEEE 802.16 mesh networks. With the millimeter wave communication emerging into 5G mobile communication systems, the millimeter wave communication has been considered the wireless backhaul solution for small cell networks. However, most studies on millimeter wave backhaul technologies focused on the design of the antenna array and radio frequency (RF) components of transceivers, such as beamforming and modulation schemes \cite{Hur2013,Dehos}. An efficient beam alignment technique using adaptive subspace sampling and hierarchical beam codebooks was proposed for implementation in small cell networks \cite{Hur2013}. The feasibility of short- and medium-distance links at millimeter wave frequencies was evaluated for wireless backhauling and the requirements on the transceiver architecture and technologies were analyzed in \cite{Dehos}.

However, in all the aforementioned ultra-dense wireless network studies, only simple scenarios, such as indoor scenarios, were considered and only basic features of 5G networks were discussed. Besides, the system level investigation of ultra-dense cellular networks with millimeter wave backhaul is lacking in the open literature. Although the distributed network architecture is recommended for ultra-dense cellular networks, the constraints and performance limits of ultra-dense cellular network employing distributed network architecture are not clear. Moreover, a key question, i.e., how dense can small cells be deployed in 5G ultra-dense cellular networks before the performance benefits fade, has not been investigated.

In this paper, we propose the distributed architecture of ultra-dense cellular network with single and multiply gateways, which can be deployed in all 5G cellular scenarios. Furthermore, based on our early proposed network capacity relationship in wireless multi-hop networks, the impact of different numbers of small cell BSs on the backhaul network capacity and the backhaul energy efficiency of ultra-dense cellular networks is investigated. Simulation results demonstrate that there exists a density threshold of small cells in ultra-dense cellular networks. When the density of ultra-dense cellular networks is larger than the density threshold, the backhaul network capacity and the backhaul energy efficiency of ultra-dense cellular networks will reduce with a further increase in small cell density. Finally, future challenges of 5G ultra-dense cellular networks are discussed, and conclusions are drawn.

\section{Architecture of 5G Ultra-Dense Cellular Networks}
With the development of massive MIMO antenna and millimeter wave communication technologies in 5G mobile communication systems, a large number of small cells will be deployed to form 5G ultra-dense cellular networks. Therefore, the first challenge is how to design the architecture of 5G ultra-dense cellular networks. In this section, the distribution architecture of ultra-dense cellular network with single and multiply gateways is proposed for further evaluating in following Sections.

\subsection{Conventional Cellular Network Architecture}
The conventional cellular network architecture is a type of tree network architecture, where every macrocell BS is controlled by the BS managers in the core network and all backhaul traffic is forwarded to the core network by the given gateway. In order to support microcells deployment, e.g., femtocells, picocells and hotspots deployment, a hybrid architecture is presented for conventional cellular networks with microcells deployment. In this hybrid network architecture, the microcell network is also configured as a type of tree network architecture, where every microcell BS is controlled by microcell BS managers in the core network and the backhaul traffic of microcell BSs is forwarded to the core network by the broadband Internet or fiber links.
The coverage of microcells is overlapped with the coverage of macrocells. Compared with macrocell BSs, microcell BSs can provide the high speed wireless transmission in indoor and hotspot scenarios. Both of the macrocell BS and the microcell BS can independently transmit the user data and the management data to associated users. Users can handover in macrocells and microcells according to their requirements. Moreover, the handover process is controlled by macrocell and microcell managers in the core network. In this network architecture, the microcell network is a complement for the conventional macrocell network to satisfy the high speed wireless transmission in partial regions, e.g., indoor and hotspot scenarios.

\subsection{Distribution Architecture of Ultra-Dense Cellular Networks}
Motivated by the massive MIMO antenna and the millimeter wave communication technologies, the densification deployment of small cells is emerging into 5G cellular networks. However, it is difficult to forward the backhaul traffic of every small cell BS by the broadband Internet or the fiber link considering the cost and geography deployment challenges in urban environments. Moreover, the small cell BS usually can not directly transmit the wireless backhaul traffic to the given gateway since small cell BSs adopting the millimeter wave technology restrict the wireless transmission distance. In this case, the wireless backhaul traffic has to be relayed to the given gateway by multi-hop links. As a consequence, the distribution network architecture is a reasonable solution for 5G ultra-dense cellular networks. In 5G ultra-dense cellular scenarios, to solve the mobile user frequently handover problem in small cells, the macrocell BS is configured only to transmit the management data for controlling the user handover in small cells and the small cell BS takes charge of the user data transmission. Therefore, the small cell network is not a complement for the macrocell network. 5G ultra-dense cellular networks is jointly composed by small cells and macrocells. Based on the backhaul gateway configuration, two distribution architectures of ultra-dense cellular networks are proposed as follows.
\subsubsection{Ultra-Dense Cellular Networks with Single Gateway}
When only one gateway is deployed in the macrocell, corresponding scenario and logical figures are illustrated in Figure 1. Without loss of generality, the gateway is configured at the macrocell BS which usually has enough space to install massive MIMO millimeter wave antennas for receiving the wireless backhaul traffic from small cells in the macrocell. The backhaul traffic of small cell BS is relayed to the adjacent small cell BS by millimeter wave links. All backhaul traffic of small cells is finally forwarded to the macrocell BS by multi-hop millimeter wave links. In the end, the backhaul traffic aggregated at the macrocell BS is forwarded to the core network by fiber to the cell (FTTC) links.
\begin{figure}[!t]
\begin{center}
\includegraphics[width=6.5in]{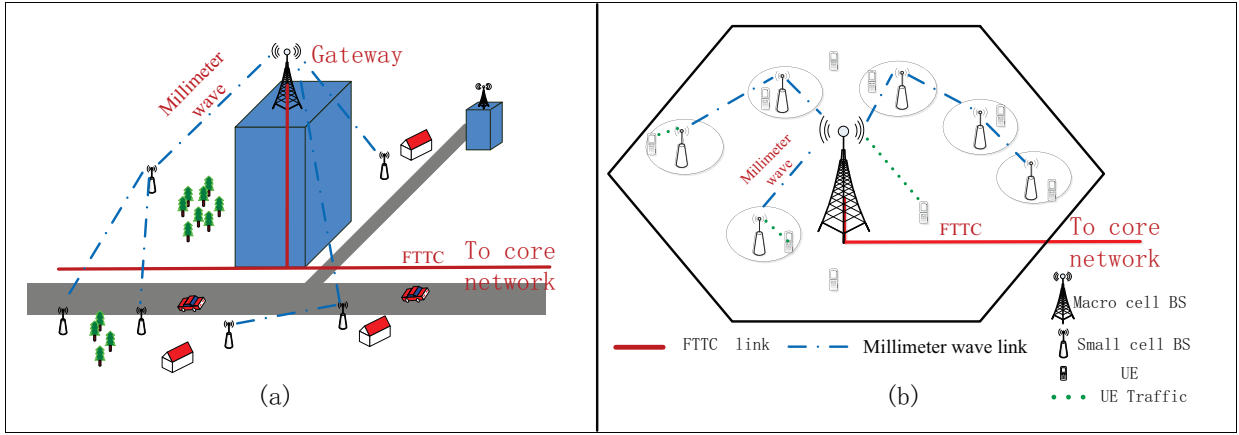}
\caption{Distribution ultra-dense cellular networks with single gateway: (a) the deployment scenario with single gateway; (b) the logical architecture with single gateway.}\label{Fig2}
\end{center}
\end{figure}

\begin{figure}[!t]
\begin{center}
\includegraphics[width=6.5in]{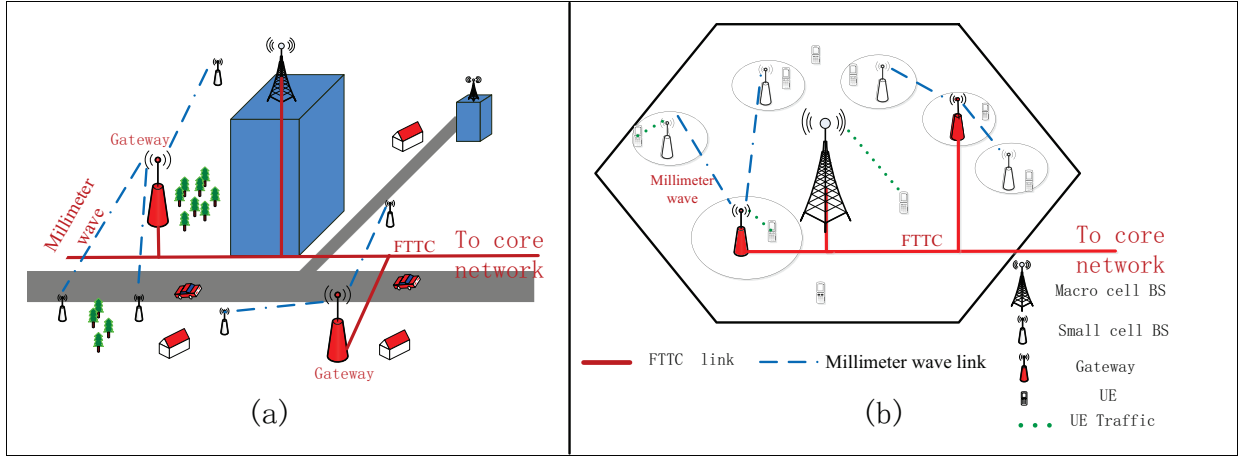}
\caption{Distribution ultra-dense cellular network with multiply gateways: (a) the deployment scenario with multiply gateways; (b) the logical architecture with multiply gateways.}\label{Fig3}
\end{center}
\end{figure}

\subsubsection{Ultra-Dense Cellular Networks with Multiply Gateways}
In the distribution architecture of ultra-dense cellular networks, the multiply gateways deployment is flexible for forwarding the backhaul traffic into the core network. In this case, gateways are deployed at multiply small cell BSs according to the requirement of backhaul traffic and geography scenarios. In Figure 2, the backhaul traffic of small cell BS is relayed to the adjacent small cell BS by millimeter wave links. Different with the single gateway configuration, the backhaul traffic of small cells will be distributed into multiply gateways in the macrocell. The backhaul traffic aggregated at the specified small cell BS, i.e., the gateway, is finally forwarded into the core network by FTTC links. Detailed scenario and logical figures are illustrated in Figure 2(a) and 2(b).

Based on comparison results in Table I, the detail differences between conventional cellular networks and 5G ultra-dense cellular networks with single/multiply gateways are explained as follows: the architecture of conventional cellular networks is a centralized network architecture and some microcells are densely deployed at partial areas, e.g. urban area, for satisfying crowed people communication requirements. When 5G small cell BSs equipped with massive MIMO antennas and millimetre wave communication technologies, the coverage of small cell has to be obviously reduced. To realize the seamless coverage, 5G cellular networks must be densely deployed by a large number of small cells. In this case, 5G ultra-dense cellular networks can provide the high-bit-rate in all cellular coverage regions. Moreover, the architecture of ultra-dense cellular networks is a distributed network architecture considering cost and geography deployment requirements. Every BS in conventional cellular networks has the same function and the coverage between macrocells and microcells is overlapped. For 5G ultra-dense cellular networks, macrocell BSs transmit the management data and small cell BSs take charge of the user data transmission. There does not exist the overlap for the function and the coverage between macrocell BSs and small cell BSs. Besides, 5G ultra-dense cellular networks with single gateway are cost efficient but the backhaul capacity bottleneck maybe exists at the single gateway. 5G ultra-dense cellular networks with multiply gateways have to spend the high cost in the small cell deployment. Compared with conventional cellular networks, 5G ultra-dense cellular networks performance will provide graceful degradation as the degree of mobility increases. To overcome this challenge, the multi-cell cooperative communication is a potential solution for 5G ultra-dense cellular networks.

\begin{table}[t!]
\centering
\normalsize{

\caption{Comparison between conventional cellular networks and 5G ultra-dense cellular networks}\label{table1}
\begin{tabular}{p{3cm}|p{4cm}|p{4cm}|p{4cm}}

\hline
\textbf{Network types}  &\textbf{Conventional cellular
networks}  &\textbf{Ultra-dense cellular networks with single gateway}   &\textbf{Ultra-dense cellular networks with multiply gateways}\\
\hline

Network architecture	&centralized architecture 	&distributed architecture \cite{Ge2014}	 &distributed architecture \\
\hline
Densification deployment target	&macrocells \cite{Yunas}	&small cells 	 &small cells \\
\hline
Densification deployment reason	&satisfy crowded people communication requirements in urban	 &massive MIMO antennas and millimeter wave communication technologies \cite{Bhushan}	 &massive MIMO antennas and millimeter wave communication technologies \\
\hline
Coverage between macrocells and microcells	&overlap \cite{Bhushan}	&not overlap	 &not overlap\\
\hline
Functions of macrocell and microcell  &same \cite{Bhushan}	&macrocells transmit management data, microcells transmit user data &macrocells transmit management data, microcells transmit user data\\
\hline
Microcells/small cells deployment	&deploy in partial areas	&deploy in all cellular scenarios 	 &deploy in all cellular scenarios\\
\hline
Backhaul method	&backhaul traffic is directly forwarded into the core network by the gateway \cite{Ge2014}	&backhaul traffic is relayed to the gateway by multi-hop wireless links	 &backhaul traffic is relayed to the gateway by multi-hop wireless links\\
\hline
Number of backhaul gateway in a macrocell	&one	&one	&multiple\\
\hline
Merit	&Flexibly deployment and low cost \cite{Wang2014}	&ubiquitous and high-bit-rate \cite{Yunas} &ubiquitous and high-bit-rate \\
\hline
Demerit	&Small cell partial deployment, low network capacity, uneven distribution of the achievable data \cite{Yunas}	&low mobility and exist the backhaul capacity bottleneck &low mobility and high cost \\
\hline
\end{tabular}
}
\vspace{0.8cm}\\
\end{table}

\section{Backhaul Network Capacity and Backhaul Energy Efficiency}
Although the density of small cells can approach the infinite in theory studies, it is unrealistic to deploy ultra-dense cellular networks with the infinite density in practical engineering applications. The impact of the deployment density of ultra-dense cellular networks on the backhaul network capacity and the backhaul energy efficiency is investigated in the following.

\subsection{Backhaul Network Capacity of Ultra-Dense Cellular Networks}
How much densification can be deployed in ultra-dense cellular networks is a key question for future 5G network designs. Utilizing massive MIMO antenna and millimeter wave communication technologies, the small cell is anticipated to provide more than 1 Gbps throughput in 5G ultra-dense cellular networks. But all small cells throughput has to be forwarded into the core network by wireless backhaul networks. Therefore, the backhaul network capacity will be a bottleneck for constraining the small cell densification in 5G ultra-dense cellular networks. The wireless multi-hop relay backhaul scheme of ultra-dense cellular networks is defined as follows:

\begin{enumerate}
\item	
 	The closest gateway is selected by the small cell BS for receiving backhaul traffic.
\item
 	 	Two conditions should be satisfied for the small cell BS which is selected for the next hop candidate: \textcircled{1} the distance between the transmitter and the receiver is less than or equal to the radius of small cell $r$; \textcircled{2} the distance between the next hop small cell BS and the gateway is less than the distance between the transmitter and the gateway; \textcircled{3} when multiple small cell BSs satisfy \textcircled{1} and \textcircled{2}, the small cell BS closing the gateway is selected as the next hop candidate;
\item
 	 	When the distance between the small cell BS and the gateway is less than $r$, the small cell BS directly transmit backhaul traffic to the gateway without relaying.
To avoid the interference from adjacent small cells, the distance of simultaneous transmission small cell BSs is configured to be larger than $\left( 1+\Delta  \right)r$, where $\Delta \times r$ is the interference protect distance in 5G ultra-dense cellular networks.
\end{enumerate}

\begin{table}[t!]
\centering
\normalsize{

\caption{Simulation parameters}\label{table2}
\begin{tabular}{p{8cm}|p{5cm}<{\centering}}

\hline
\textbf{Parameters}  &\textbf{Values}  \\
\hline
Number of backhaul gateways in a macrocell	&3\\
\hline
Radius of small cell $r$	&100m, 150m, 200m\\
\hline
Radius of macrocell	&1km\\
\hline
Parameter $a$  &7.85\\
\hline
Parameter $b$  &71.5Watt\\
\hline
Normalized BS backhaul transmission power ${P_{Norm}}$	&1Watt\\
\hline
Normalized BS backhaul throughput $T{h_0}$ 	&1Gbps\\
\hline
Lifetime of small cell BS ${T_{Lifetime}}$	&5years\\
\hline
Embodied energy consumption ${E_{EM}}$	&20\% of total energy consumption\\

\hline
\end{tabular}
}
\vspace{0.8cm}\\
\end{table}

Based on our early results in \cite{Mao}, a simple relationship is proposed to estimate the backhaul network capacity of ultra-dense cellular networks as follows:\[\text{Backhaul network capacity =}\frac{Y\left( n \right)\times W}{k\left( n \right)},\] where $n$ denotes the number of small cell BSs in a macrocell, $Y\left( n \right)$ is the average number of simultaneous transmissions in the macrocell, $W$ is the transmission rate of small cell BS, $k\left( n \right)$ is the average hop number of wireless backhaul traffic in the macrocell.
Without loss of generality, the 5G ultra-dense cellular network with multiple gateways shown in Figure 2 is considered for the following simulation analysis. The macrocell is assumed to be a regular hexagon with 1 km radius. Small cell BSs are scattered following a Poisson point process in a macrocell. All small cells are assumed not to overlap each other in the coverage. Moreover, three gateways are assumed to be symmetrically deployed at top vertices of the hexagon macrocell. The interference safeguard distance is configured as $0.5 \times r$ and the transmission rate of small cell BS is normalized as 1 Gbps in the following simulations. The detailed simulation parameters are configured in Table II.

Based on the Monte-Carlo simulation method, the backhaul network capacity and the backhaul energy efficiency of ultra-dense cellular networks are simulated in Figure 3 and Figure 4, respectively. When the radius of small cell $r$ is fixed, the backhaul network capacity with respect to the number of small cell BSs is illustrated in Figure 3(a): the backhaul network capacity first increases with the increase of the number of small cell BSs; after the backhaul network capacity achieves the maximum threshold, the backhaul network capacity decreases with the increase of the number of small cell BSs; in the end, the backhaul network capacity achieves a stationary saturation value when the number of small cell BS approaches to the infinite. When the radius of small cell $r$ is fixed, the backhaul network capacity with respect to the average number of simultaneous transmissions is described in Figure 3(b): considering the interference protect distance $\Delta  \times r$ configured by the wireless multi-hop relay backhaul scheme, the maximum average number of simultaneous transmissions decreases with the increase of the radius of small cell when the macrocell coverage is fixed. For example, the maximum average number of simultaneous transmissions is 29, 25 and 19 when the radius of small cell is configured as 100 m, 150 m and 200 m, respectively. The backhaul network capacity increases with the increase of the average number of simultaneous transmissions in the macrocell. Moreover, the backhaul network capacity approaches to a saturation limit when the average number of simultaneous transmissions is larger than 27, 23 and 15 which correspond to the radius of small cell 100 m, 150 m and 200 m, respectively. When the number of small cell BSs or the average number of simultaneous transmissions is fixed, the backhaul network capacity decreases with increase of the radius of small cells. Based on simulation results in Figure 3(a), the backhaul network capacity will achieve a stationary saturation value when the average number of simultaneous transmissions or the dense of small cell BSs, i.e., the number of small cell BSs in a macrocell is larger than a given threshold. This result provide a guideline for designing the densification of 5G ultra-dense cellular networks.

\begin{figure}[!t]
\begin{center}
\includegraphics[width=6.5in]{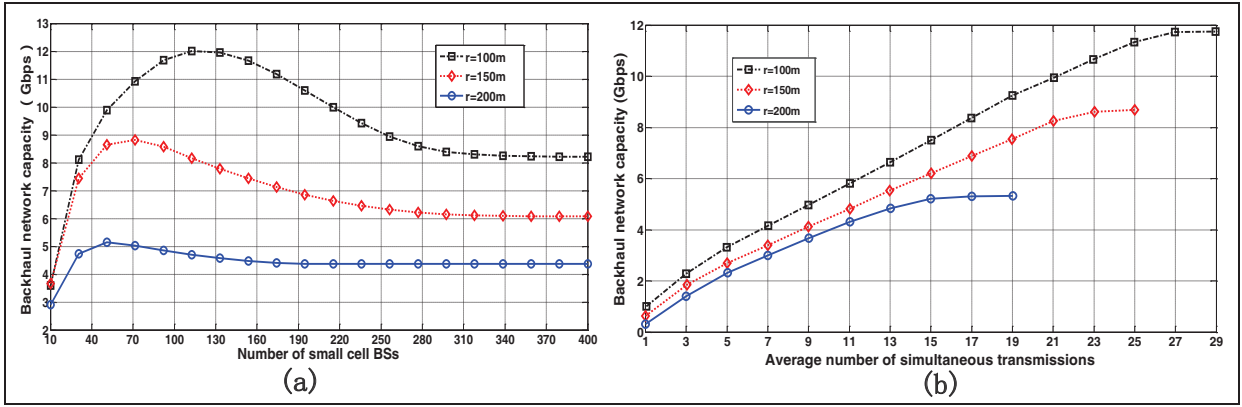}
\caption{Backhaul network capacity of ultra-dense cellular networks: (a) the backhaul network capacity vs the number of small cell BSs; (b) the backhaul network capacity vs the average number of simultaneous transmissions.}\label{Fig4}
\end{center}
\end{figure}

\subsection{Backhaul Energy Efficiency of Ultra-Dense Cellular Networks}
Expect for the backhaul network capacity, the backhaul energy efficiency is another key constrain parameter which restrict the densification of 5G ultra-dense cellular networks. The backhaul energy consumed at the small cell BS is decomposed by the embodied energy ${{E}_{\text{EM}}}$ and the operation energy ${{E}_{\text{OP}}}$ \cite{Humar}. The embodied energy is the energy consumed by all processes associated with the BS production and is accounted for the 20\% of the backhaul BS energy consumption in this paper. The operation energy is the energy consumed for the backhaul operation in the lifetime ${{T}_{\text{Lifetime}}}$ and is defined by ${{E}_{\text{OP}}}={{P}_{\text{OP}}}\times {{T}_{\text{Lifetime}}}$, where ${{P}_{\text{OP}}}$ is the BS operating power. Without loss of generality, the small cell BS operating power is assumed as the linear function of the small cell BS backhaul transmission power ${{P}_{\text{TX}}}$ and is expressed as ${{P}_{\text{OP}}}=a\times {{P}_{\text{TX}}}+b$, where $a=7.84$ and $b=71.5$ Watt \cite{Hoydis2011}. In general, the BS backhaul transmission power depends on the BS backhaul throughput. To simplify the model derivation, the backhaul transmission power of small cell BS is normalized as ${{P}_{\text{Norm}}}=1$ Watt when the normalization BS backhaul throughput $T{{h}_{0}}$ is assumed as 1 Gbps. Similarly, the small cell BS backhaul transmission power with the average BS backhaul throughput $T{{h}_{\text{Avg}}}$ is denoted by ${{P}_{\text{TX}}}={{P}_{\text{Norm}}}\times \left( {T{{h}_{\text{Avg}}}}/{T{{h}_{0}}}\; \right)$, where the average small cell BS backhaul throughput is calculated by the backhaul network capacity \cite{Mao}. Furthermore, the small cell BS operating power is calculated by ${{P}_{\text{OP}}}=a\times {{P}_{\text{Norm}}}\times \left( {T{{h}_{\text{Avg}}}}/{T{{h}_{0}}}\; \right)+b$. In the end, the backhaul energy efficiency of ultra-dense cellular networks is derived by \[\text{Backhaul energy efficiency =}\frac{\text{backhaul network capacity}}{n\times \left( \text{small cell BS backhaul energy consumption} \right)}.\]
Without loss of generality, the lifetime of small cell BS is configured as ${{T}_{\text{Lifetime}}}=5$ years. When the radius of small cell $r$ is fixed, the backhaul energy efficiency of ultra-dense cellular networks with respect to the number of small cell BSs is analyzed in Figure 4(a): the backhaul energy efficiency first increases with the increase of the number of small cell BSs; and then, the backhaul energy efficiency decreases with the increase of the number of small cell BSs after the backhaul energy efficiency comes up to the maximum threshold; in the end, the backhaul energy efficiency of ultra-dense cellular networks achieves to a stationary saturation value when the number of small cell BSs approaches to the infinite. When the number of small cell BSs is fixed, the backhaul energy efficiency increases with the increase of the small cell radius when the number of small cell BSs is less than 10. When the number of small cell BSs is larger than or equal to 10, the backhaul energy efficiency decreases with the increase of the small cell radius. When the radius of small cell $r$ is fixed, the backhaul energy efficiency with respect to the average small cell BS throughput is illustrated in Figure 4(b): the backhaul energy efficiency first increases with the increase of the average small cell BS throughput; and then, the backhaul energy efficiency decreases with the increase of the average small cell BS throughput after the backhaul energy efficiency achieves the maximum threshold; in the end, the backhaul energy efficiency of ultra-dense cellular networks achieves to a stationary saturation value when the average small cell BS throughput is larger than 0.35, 0.45 and 0.5 Gbps which correspond to the radius of small cell 200, 150, 100 meters.

\begin{figure}[!t]
\begin{center}
\includegraphics[width=6.5in]{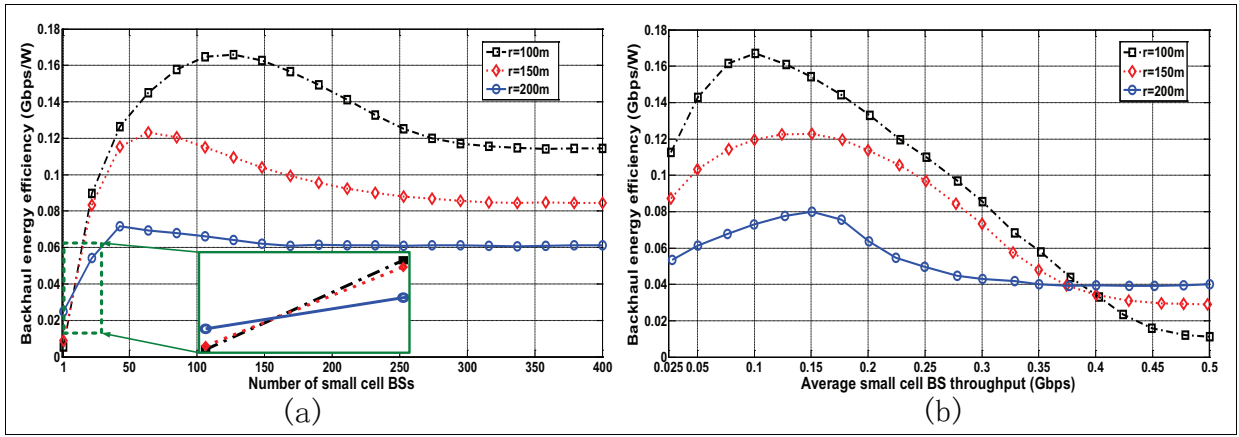}
\caption{Energy efficiency of ultra-dense cellular networks: (a) backhaul energy efficiency vs the number of small cell BSs; (b) backhaul energy efficiency vs average small cell BS throughput.}\label{Fig5}
\end{center}
\end{figure}

\section{Future Challenges}
As we discussed in the above sections, the emergence of ultra-dense cellular network is motived by massive MIMO antenna and millimeter wave communication technologies. Moreover, the distribution network architecture is a reasonable solution for 5G ultra-dense cellular networks. Compared with results in Table I, it is obvious that the ultra-dense cellular network would bring great changes into future 5G cellular networks. Therefore, the ultra-dense cellular network is one of the most important challenges for future 5G cellular networks. Some potential challenges are presented in the following context.

The first challenge is the multi-hop relay optimization in 5G ultra-dense cellular networks. In the distribution network architecture, not only backhaul traffic but also fronthaul traffic needs to be relayed into the destination. The selection of relaying small cell BS should be carefully considered in 5G ultra-dense cellular networks. Hence, the wireless multi-hop routing algorithm is a key challenge for 5G ultra-dense cellular networks. Although the small cell BS equipped with massive MIMO antennas has enough antennas for simultaneously transmitting backhaul traffic and fronthaul traffic, it is another important challenge how to reasonably allocate massive antennas for backhaul and fronthaul transmissions.
The small cell coverage of ultra-dense cellular networks is obviously less than the macrocell coverage of conventional cellular networks. For a high-speed mobile user, the user frequently handover in small cells not only increase redundant overhead but also decrease the user experience. Moreover, the wireless transmission of small cell BS equipped with millimeter wave antennas and beamforming technologies has strong directivity, which is to the advantage of high-speed transmission and the disadvantage of covering the high-speed mobile user. The cooperative transmission of small cells is a potential solution for this problem. How to organize adjacent small cells for cooperative transmission is the second challenge for 5G ultra-dense cellular networks. For example, how to dynamically group small cells for seamlessly covering the high-speed mobile user track is an open issue.
With the emergence of millimeter wave communication technology for 5G wireless transmission, the beamforming method will be widely used. When the beamforming method is performed by massive MIMO antennas, the computation scale of beamforming method and the computation power of wireless transceivers will be obviously increased by the large scale of signal processing in BS baseband processing systems. Therefore, the proportion between the computation power and transmission power maybe reversed at wireless transceivers adopting massive MIMO antenna and millimeter wave communication technologies. In this case, the computation power can not be ignored for the BS energy consumption. Considering the proportion change between the computation power and the transmission power, the new energy efficiency model need to be investigated for ultra-dense cellular networks with massive MIMO antenna and millimeter wave communication technologies. To face with above challenges in 5G ultra-dense cellular networks, some potential research directions are summarized to solve these issues:
\begin{enumerate}
\item	
    The new multi-hop relay scheme and the distribution routing algorithm should be developed for 5G ultra-dense cellular networks.
\item
  	Massive MIMO antennas and millimeter wave communications provide enough resource space for small cell BSs. How to utilize and optimize the resource allocation for BS relaying and self-transmission is a critical problem in 5G ultra-dense cellular networks.
\item
 	The cooperative transmission and backhaul transmission will become two of important directions in future 5G ultra-dense cellular networks.
\item
 	Motived by massive MIMO antenna and millimeter wave communication technologies, the computation power consumed for BS baseband processing systems need to be rethought for 5G ultra-dense cellular networks.
\end{enumerate}

\section{Conclusions}
Until recently, ultra-dense wireless networks have been mainly deployed in parts of network areas, such as indoor and hotspot scenarios. Ultra-dense wireless network are still considered as a complement for cellular networks with centralized network architecture. The massive MIMO antennas and millimeter wave communication technologies enable 5G ultra-dense cellular networks to be deployed in all cellular scenarios. In this paper, the distributed network architecture with single and multiply gateways are presented for 5G ultra-dense cellular networks. Considering the millimeter wave communication technology, the impact of small cell BS density on the backhaul network capacity and the backhaul energy efficiency of ultra-dense cellular networks is investigated. Simulation results indicate that there exists a density threshold of small cells in ultra-dense cellular networks. When the density of ultra-dense cellular networks is larger than the density threshold, the backhaul network capacity and the backhaul energy efficiency of ultra-dense cellular networks will reduce with a further increase in small cell density. These results provide some guidelines for the optimum deployment of 5G ultra-dense cellular networks.

In 2G and 3G mobile communication systems, the wireless communication system has been considered as a noised-limited communication system. With the MIMO antenna technology being adopted in 4G mobile communication systems, the wireless communication system has been transited into an interference-limited communication system. In this paper, it has been shown that there exists a maximum backhaul network capacity corresponding to a given number of small cell BSs in a macrocell, termed by us as the density threshold of ultra-dense cellular network. When the density of ultra-dense cellular networks, measured by the number of small cells per macro-cell, is larger than the density threshold, the backhaul network capacity will reduce with a further increase in the density. Moreover, a similar bottleneck is also observed in the backhaul energy efficiency of ultra-dense cellular networks. As a consequence, we conclude that the 5G ultra-dense cellular network is a density-limited communication system. How to analytically determine the optimum density of small cell BSs in 5G ultra-dense cellular networks is an open issue. If this is done, a veritable challenge would indeed emerge in the next round of the telecommunications revolution.

\section*{Acknowledgments}

The corresponding author of the article is Prof. Tao Han. The authors
would like to acknowledge the support from the International Science
and Technology Cooperation Program of China (Grant No. 2014DFA11640 and 2015DFG12580), the National Natural Science Foundation of
China (NSFC) (Grant No. 61271224, 61301128 and 61471180), the NSFC
Major International Joint Research Project (Grant No. 61210002), the China 863 Project in 5G Wireless Networking (Grant No. 2014AA01A701), the
Hubei Provincial Science and Technology Department (Grant No. 2013BHE005),
the Fundamental Research Funds for the Central Universities (Grant
No. 2015XJGH011 and 2014QN155), EU FP7-PEOPLE-IRSES
(Contract/Grant No. 247083, 318992, 612652 and 610524), and EU H2020 ITN 5G Wireless project (Grant No. 641985). This research is also supported by Australian Research Council Discovery
projects DP110100538 and DP120102030.

\begin{IEEEbiographynophoto}{Xiaohu Ge}
 {[}M'09-SM'11{]} (xhge@hust.edu.cn)  is currently a full professor with the School of Electronic Information and Communications at Huazhong University of Science and Technology (HUST) China and an adjunct professor with the Faculty of Engineering and Information Technology at University of Technology Sydney (UTS), Australia. He received his Ph.D. degree in communication and information engineering from HUST in 2003. He serves as an Associate Editor for \emph{IEEE Access}, the \emph{Wireless Communications and Mobile Computing Journal}, and so on.
\end{IEEEbiographynophoto}

\begin{IEEEbiographynophoto}{Song Tu}
 (u201013039@hust.edu.cn) received his B.E. degrees from HUST, China, in 2014. Now he continues to study for a master’s degree in the School of Electronic Information and Communications at the HUST. His research interests are in the area of green communications and distributed wireless networks.
\end{IEEEbiographynophoto}

\begin{IEEEbiographynophoto}{Guoqiang Mao}
 {[}S'98-M'02-SM'08{]} (g.mao@ieee.org)  is a Professor of Wireless Networking, Director of Center for Real-time Information Networks at the UTS. He has published more than 100 papers in international conferences and journals, which have been cited more than 3000 times.
\end{IEEEbiographynophoto}

\begin{IEEEbiographynophoto}{Cheng-Xiang Wang}
 {[}S'01-M'05-SM'08{]}  (cheng-xiang.wang@hw.ac.uk) received his Ph.D. degree from Aalborg University, Denmark, in 2004. He has been with Heriot-Watt University since 2005 and became a professor in 2011. His research interests include wireless channel modelling and 5G wireless communication networks. He has served or is serving as an Editor or Guest Editor for 11 international journals, including \emph{IEEE Transactions on Vehicular Technology} (2011-), \emph{IEEE Transactions on Wireless Communications} (2007-2009), and \emph{IEEE Journal on Selected Areas in Communications}. He has published one book and over 210 papers in journals and conferences.
\end{IEEEbiographynophoto}

\begin{IEEEbiographynophoto}{Tao Han}
 {[}M'13{]} (hantao@hust.edu.cn) received the Ph.D. degree in Communication
and Information Engineering from HUST, Wuhan, China in December, 2001. He is currently
an Associate Professor with the School of Electronic Information and Communications, HUST. His research interests include wireless communications,
multimedia communications, and computer networks.
\end{IEEEbiographynophoto}


\begin{thebibliography}{1}

\bibitem{Hoydis2013}
J. Hoydis, S. Ten Brink and M. Debbah,  ``Massive MIMO in the UL/DL of Cellular Networks: How Many Antennas Do We Need?" \emph{IEEE J. Sel. Areas Commun.}, Vol. 31, No. 2, pp. 160--171, Feb. 2013.

\bibitem{Rappaport}
T.S. Rappaport, S. Shu, R. Mayzus, et al., ``Millimeter Wave Mobile Communications for 5G Cellular: It Will Work!" \emph{IEEE Access}, Vol. 1, pp. 335--349, May 2013.

\bibitem{Wang2014}
C.-X. Wang, F. Haider, X. Gao, et al., ``Cellular Architecture and Key Technologies for 5G Wireless Communication Networks," \emph{IEEE Commun. Mag.}, Vol. 52, No. 2, pp. 122--130, Feb. 2014.

\bibitem{Yunas}
S. F. Yunas, M. Valkama and J. Niemela, ``Spectral and Energy Efficiency of Ultra-Dense Networks under Different Deployment Strategies," \emph{IEEE Commun. Mag.}, Vol. 53, Vol. 1, pp. 90--100, Jan. 2015.

\bibitem{Soret}
B. Soret, K. I. Pedersen, N. T. K. Jorgensen and V. F. Lopez, ``Interference Coordination for Dense Wireless Networks," \emph{IEEE Commun. Mag.}, Vol. 53, Vol. 1, pp. 102--109, Jan. 2015.

\bibitem{Asadi}
A. Asadi, V. Sciancalepore and V. Mancuso, ``On the Efficient Utilization of Radio Resource in Extremely Dense Wireless Networks," \emph{IEEE Commun. Mag.}, Vol. 53, Vol. 1, pp. 126--132, Jan. 2015.

\bibitem{Bhushan}
N. Bhushan, J. Li, D. Malladi, et al., ``Network Densification: The Dominant Theme for Wireless Evolution in 5G," \emph{IEEE Commun. Mag.}, Vol. 52, Vol. 2, pp. 82--89, Feb. 2014.

\bibitem{Condoluci}
M. Condoluci, M. Dohler, G. Araniti, et al., ``Toward 5G DenseNets: Architectural Advances for Effective Machine-Type Communications over Femtocell," \emph{IEEE Commun. Mag.}, Vol. 53, Vol. 1, pp. 134--141, Jan. 2015.

\bibitem{Ge2014}
X. Ge, H. Cheng, M. Guizani, T. Han, ``5G Wireless Backhaul Networks: Challenges and Research Advances," \emph{IEEE Netw.}, Vol. 28, No. 6, pp. 6--11, Nov. 2014.

\bibitem{He2007}
J. He, K. Yang, K. Guild and H.-H. Chen, ``Application of IEEE 802.16 Mesh Networks as the Backhaul of Multihop Cellular Networks," \emph{IEEE Commun. Mag.}, Vol. 45, No. 9, pp. 82-90, Sept. 2007.

\bibitem{Hur2013}
S. Hur, T. Kim, D. J. Love, J. V. Krogmeier, T. A. Thomas and A. Ghosh, ``Millimeter Wave Beamforming for Wireless Bachaul and Access in Small Cell Networks," \emph{IEEE Trans. Wireless Commun.}, Vol. 61, No. 10, pp. 4391-4403, Oct. 2013.

\bibitem{Dehos}
C. Dehos, J. L. Gonzalez, A. D. Dmoenico, D. Ktennas and L. Dussopt, ``Millimeter-Wave Access and Backhauling: The Solution to the Exponential Data Traffic Increase in 5G Mobile Communications Systems?" \emph{IEEE Commun. Mag.}, Vol. 52, Vol. 9, pp. 88-95, Sept. 2014.

\bibitem{Mao}
G. Mao, Z. Lin, X. Ge and Y. Yang, ``Towards a Simple Relationship to Estimate the Capacity of Static and Mobile Wireless Networks," \emph{IEEE Trans. Wireless Commun.}, Vol. 12, No. 8, pp. 3883--3895, Aug. 2013.

\bibitem{Humar}
I. Humar, X. Ge, L. Xiang, J. Ho, M. Chen, ``Rethinking Energy-Efficiency Models of Cellular Networks with Embodied Energy," \emph{IEEE Netw.}, Vol.25, No.3, pp.40--49, March 2011.

\bibitem{Hoydis2011}
J. Hoydis, M. Kobayashi and M. Debbah, ``Green Small-Cell Networks," \emph{IEEE Vehic. Tech. Mag.}, Vol. 6, No. 1, pp. 37--43, March 2011.


\end{thebibliography}
\end{document}